# Accommodating heterogeneous missing data patterns for prostate cancer risk prediction

Matthias Neumair[1], Michael W. Kattan[2], Stephen J. Freedland[3,4], Alexander Haese[5], Lourdes Guerrios-Rivera[6], Amanda M. De Hoedt[3], Michael A. Liss[7], Robin J. Leach[8], Stephen A. Boorjian[9], Matthew R. Cooperberg[10], Cedric Poyet[11], Karim Saba[11], Kathleen Herkommer[12], Valentin H. Meissner[12], Andrew J. Vickers[13], Donna P. Ankerst[1,14]

[1] Department of Life Sciences, Technical University of Munich, Freising, Germany
[2] Department of Quantitative Health Sciences, Cleveland Clinic Foundation, Cleveland, Ohio, USA
[3] Department of Urology, Durham Veterans Administration Health Care System, Durham, North Carolina, USA
[4] Department of Surgery, Cedars-Sinai Medical Center, Los Angeles, California, USA
[5] Martini-Clinic Prostate Cancer Center, University Clinic Eppendorf, Hamburg, Germany
[6] Department of Surgery, Urology Section, Veterans Affairs Caribbean Healthcare System, San Juan, Puerto Rico
[7] Department of Urology, University of Texas Health at San Antonio, San Antonio, Texas, USA
[8] Department of Cell Systems and Anatomy, University of Texas Health at San Antonio, San Antonio, Texas, USA
[9] Department of Urology, Mayo Clinic, Rochester, Minnesota, USA
[10] Departments of Urology and Epidemiology & Biostatistics, University of California San Francisco, San Francisco, California, USA
[11] Department of Urology, University Hospital of Zurich, University of Zurich, Zurich, Switzerland
[12] Department of Urology, University Hospital, Technical University of Munich, Germany
[13] Department of Epidemiology & Biostatistics, Memorial Sloan Kettering Cancer Center, New York, New York, USA
[14] Department of Mathematics, Technical University of Munich, Boltzmannstrasse 3, Garching, Germany

Funding was provided by grants CA179115, P50-CA92629, P30-CA008748, W81XWH-15-1-0441, P30 CA054174, K24 – CA160653. The contents of this Publication do not represent the views of the VA Caribbean Healthcare System, the Department of Veterans Affairs or the US Government. Opinions, interpretations, conclusions and recommendations are those of the author and are not necessarily endorsed by the US Department of Defense.

# ABSTRACT

**Objective:** We compared six commonly used logistic regression methods for accommodating missing risk factor data from multiple heterogeneous cohorts, in which some cohorts do not collect some risk factors at all, and developed an online risk prediction tool that accommodates missing risk factors from the end-user.

**Study Design and Setting:** Ten North American and European cohorts from the Prostate Biopsy Collaborative Group (PBCG) were used for fitting a risk prediction tool for clinically significant prostate cancer, defined as Gleason grade group $\geq 2$ on standard TRUS prostate biopsy. One large European PBCG cohort was withheld for external validation, where calibration-in-the-large (CIL), calibration curves, and area-underneath-the-receiver-operating characteristic curve (AUC) were evaluated. Ten-fold leave-one-cohort-internal validation further validated the optimal missing data approach.

**Results:** Among 12,703 biopsies from 10 training cohorts, 3,597 (28%) had clinically significant prostate cancer, compared to 1,757 of 5,540 (32%) in the external validation cohort. In external validation, the available cases method that pooled individual patient data containing all risk factors input by an end-user had best CIL, under-predicting risks as percentages by 2.9% on average, and obtained an AUC of 75.7%. Imputation had the worst CIL (-13.3%). The available cases method was further validated as optimal in internal cross-validation and thus used for development of an online risk tool. For end-users of the risk tool, two risk factors were mandatory: serum prostate-specific antigen (PSA) and age, and ten were optional: digital rectal exam, prostate volume, prior negative biopsy, 5-alpha-reductase-inhibitor use, prior PSA screen, African ancestry, Hispanic ethnicity, first-degree prostate-, breast-, and second-degree prostate-cancer family history.

# 1. Introduction

The Prostate Biopsy Collaborative Group (PBCG) was established with the aim to improve the understanding of heterogeneity in prostate cancer biopsy outcomes across international clinical centers [1]. Fig 1 shows the range of number of biopsies and prevalence of clinically significant prostate cancer, defined as Gleason grade group $\geq 2$, across 11 PBCG cohorts. Previously, the PBCG developed an online risk tool based on the small set of standard risk factors routinely collected in practice: prostate-specific antigen (PSA), digital rectal exam (DRE), age, African ancestry, first-degree prostate cancer family history, and history of a prior negative prostate biopsy [2]. For developing the prior tool, multiple methods for aggregating clinical data on a small number of variables across heterogeneous centers comprising different risk factor distributions and risk factor-outcome associations were compared. The simplest approach of pooling individual-level data and fitting a multiple logistic regression model proved to be most accurate [3]. The resulting risk calculator was published online at riskcalc.org to facilitate its use in daily routine. [4-8].

The PBCG had requested additional risk factors to those included in the current tool from its participating cohorts, but these were less rigorously collected, with some cohorts not collecting some of the risk factors at all (Fig 2). We wanted to develop an adaptive tool using all the information available in Fig 2 that would allow the user to enter as much (or as little) information as possible.

**Figure 1.** Sample sizes of 11 PBCG cohorts ranked by clinically significant prostate cancer prevalence. The 3rd cohort in black outline was withheld to serve as an external validation cohort with the remaining 10 cohorts used for training prediction models.

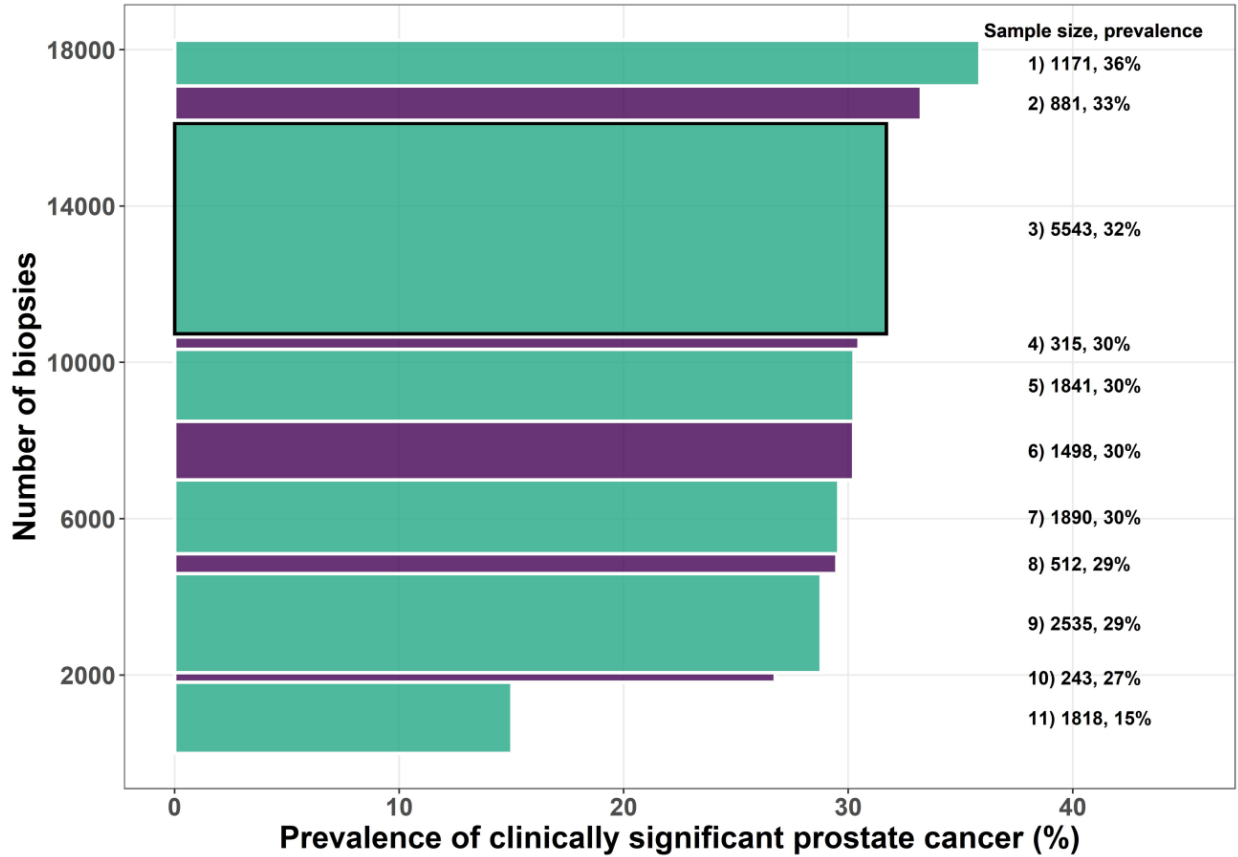

Missing data in clinical research is a ubiquitous problem, and a large number of statistical methods to account for it have been proposed [9, 10]. Most methods are applied to missing values in training data sets used to develop a model, but with the emerging use of online and electronic record embedded clinical risk tools, approaches for handling missing risk factors on the user end of a risk tool requiring the predictor are coming into play. Recently, real-time imputation was proposed to extend needed cardiovascular disease management for patients with missing risk factors [11].

The aim of this study was to construct a clinically significant prostate cancer risk tool that would optimize the use of data from heterogeneous cohorts with varying missing data patterns and allow end-users of the tools access even when missing some risk factors. In terms of development of a risk model on multiple cohorts with varying missing data patterns, we found four philosophically distinct approaches: available case analyses, ensembles of cohort-specific models, missing indicator methods, and imputation. We compared six variations of these approaches and selected an optimal one for this application. For the end-user side, we adopted an individual patient tailored approach as we have implemented in previous tools, whereby the user inputs the risk factors he has available and a resulting prediction based on those risk factors is returned [2, 12].

**Figure 2.** Missing risk factors. Amount of missing risk factor data by cohort on the x-axis; all patients were required to have prostate-specific antigen (PSA) and age, hence 0% missing for these covariates. The 3rd cohort separated by the black vertical line is used as an external validation set, and leave-one-cohort-out cross-validation was applied to the other cohorts. Cohorts were sorted by missing data pattern.

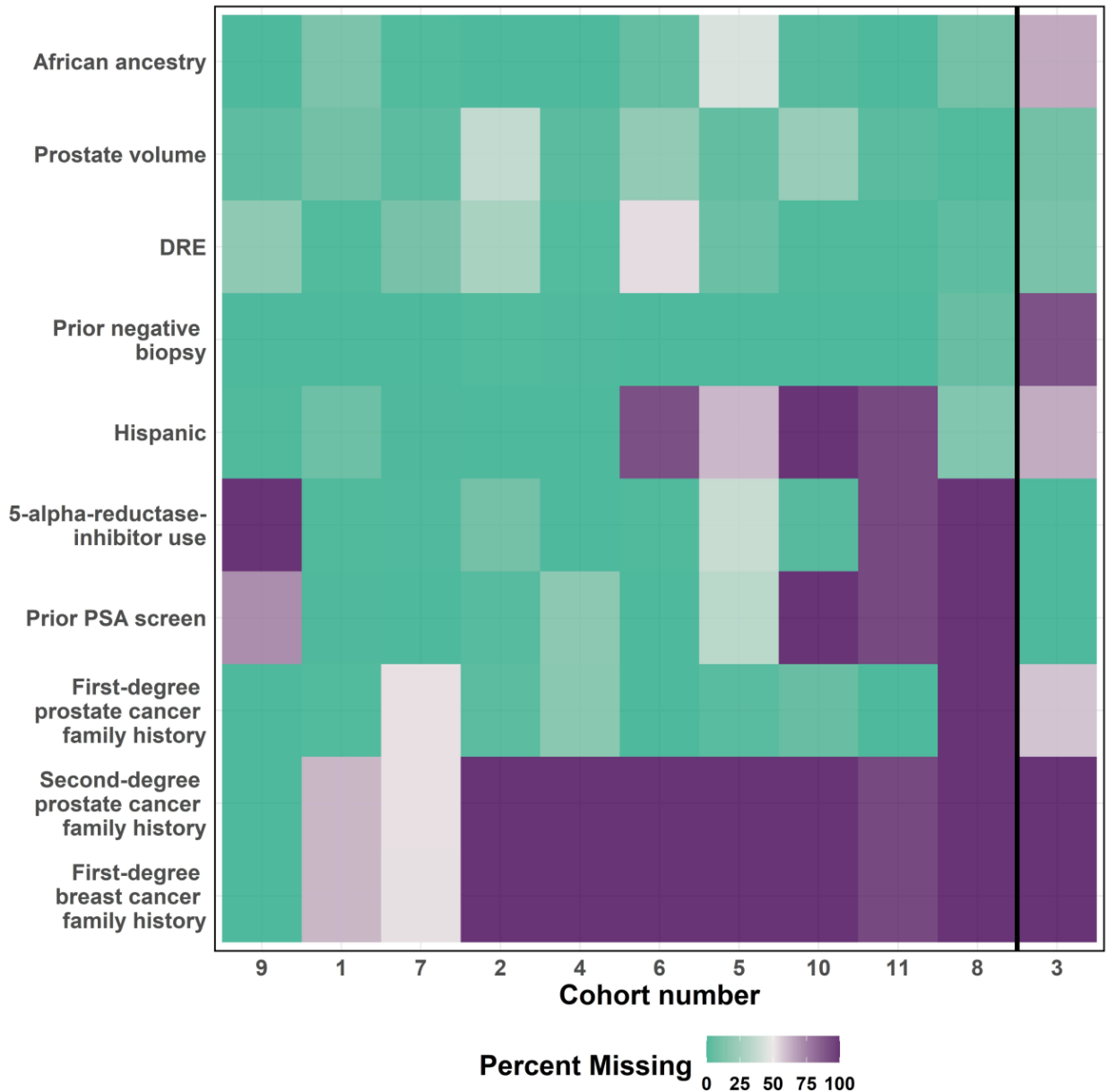

## 2. Methods

The study was based on risk factor and outcome data collected from January 2006 to December 2019 from trans-rectal systematic 10-12 core biopsies from 10 PBCG cohorts spanning North America and Europe used for training and one PBCG European cohort used for validation (Figs 1, 2, S1). The risk factors collected included the standard risk factors used in clinical practice for prostate cancer diagnosis along with other less commonly used risk factors, but with proven associations to prostate cancer. All PBCG data were collected following local institutional review board (IRB) approval from the University of Texas Health Science Center of San Antonio, Memorial Sloan Kettering Cancer Center (MSKCC), Mayo Clinic, University of California San Francisco, Hamburg-Eppendorf University Clinic, Cleveland Clinic, Sunnybrook Health Sciences Centre, VA Caribbean Healthcare System, VA Durham, San Raffaele Hospital, and University Hospital Zurich. PBCG data and analyses for this study were approved by the Technical University of Munich Rechts der Isar Hospital. As data collected were anonymized and obtained as part of standard clinical care, consent was waived by all IRB's, except regarding second-degree prostate cancer and first-degree breast cancer family history for the VA Durham. Written consent for these variables was obtained and documented as part of a larger separate study at the VA Durham prior to the beginning of this study. All institutional PBCG IRB approvals are maintained by the MSKCC central data coordinating center and IRB.

The 10 cohorts used for training the model followed the PBCG prospective protocol in data collection, whereas the external validation cohort supplied retrospective data from a single institution that performs a high annual number of prostate biopsies to the PBCG [2, 3]. Included data came from patients who had received a prostate biopsy following a PSA test under local standard-of-care and may be seen as representative of patients in North America, including Puerto Rico, and Europe. MRI biopsies as well as prostate biopsies from patients with prostate cancer were excluded. Clinically significant prostate cancer was defined as Gleason grade group $\geq 2$ on biopsy [13]. For users of the developed risk calculator, two risk factors were mandatory: PSA and age. Ten risk factors were optional: DRE, prostate volume, prior negative biopsy, 5-alpha-reductase-inhibitor use, prior PSA screen (yes/no), African ancestry, Hispanic ethnicity, first- and second-degree prostate cancer- and first-degree breast cancer-family history.

We performed a literature search to identify the six most commonly used approaches for handling missing data in multivariable logistic regression modeling, for either single or multiple cohorts as found in this study. All of the approaches could be implemented in the R statistical package. Our aim was to identify the most accurate approach for implementation in the online tool. To increase flexibility of the tool, we tailored each method to the specific list of risk factors available for an individual. That is, for a validation set, the algorithms were applied for each individual in the validation set separately. All algorithms return logistic-regression-based expressions for probability of clinically significant prostate cancer; the cohort ensemble approach averages these for the individual cohorts. The methods are summarized in Tables 1 and S1.

**Table 1.** Methods for fitting individual predictor-specific risk models for members of a test set by combining data from multiple cohorts. All individuals in the training and test cohorts have 2 predictors, PSA and age, and then any subset, including none, of 10 additional predictors for a total of 12 predictors, denoted by $X$. The set of predictors available for the new individual is denoted by $X^*$. All models use logistic regression for prediction of clinically significant prostate cancer. MICE=Multiple imputation by chained equations; BIC = Bayesian Information Criterion defined as the log likelihood – (number of covariates) × log(sample size).

| Method | Definition |
| --- | --- |
| Available cases | Pool individual-level data that have $X^*$ measured across all cohorts and fit a model including $X^*$ as main effects. |
| Iterative BIC selection | Same as available cases, but with an iterative stepwise BIC-based model selection to determine the optimal subset of $X^*$ and interactions. |
| Cohort ensemble | Separate models are built to each cohort by using the coinciding variables of the cohort and the patient. |
| Categorization | All individuals in all cohorts are used. Predictors are categorized with missing as one of the categories so that the complete list of predictors $X$ is used. |
| Missing indicator | Include an indicator for missing a continuous predictor value and the interaction with the predictor as additional variables in the analysis. Mostly similar to Categorization. |
| Imputation | Impute missing covariates in the training set following the MICE method. Mean imputation for missing values in prediction. |

The available cases algorithm pooled individual level data from the training cohorts with information on the variables that the end-user had available, fit a main effects logistic regression model for clinically significant prostate cancer to the training data, and used the coefficients in a tailored prediction model for the target patient. The iterative Bayesian information criterion (BIC) selection method added stepwise BIC-based model selection to the available cases algorithm, allowing two-way interactions to be included. If a risk factor was not chosen in the optimal model by the selection process, the procedure was re-started excluding the risk factor, allowing for a greater number of individuals from the training set to be included in model development.

Rather than pooling data across cohorts, the cohort ensemble method constructed separate models for each cohort, restricting to risk factors available by the end-user and collected by the training cohort. A risk factor was considered available in a training cohort if it was measured in 40% or more participants, otherwise it was considered missing and not included so as not to prohibitively reduce the sample size for constructing a cohort-specific model. Because models were fit to single cohorts and some of the cohorts had small sample sizes, information from individual cohorts could be low or considered inadequate for robust multivariable model construction, as for example, cohort 10 with only 243 biopsies. Such cohorts were not excluded because while they may lack power for obtaining statistical significance of individual coefficients, the goal here was optimizing out-of-sample prediction. Cohort-specific risks were averaged over the cohorts for the result provided to the end-user.

The categorization algorithm returned to pooling data across all training cohorts, and additionally transformed all continuous risk factors to categorical so that missing could be added as an extra category. For inherently categorical risk factors, such as DRE, categories were coded as normal, abnormal, and missing. Prostate volume was stratified to < 30, 30 - 50 and > 50 cc, as previously suggested so that it could be obtained by pre-biopsy DRE or TRUS, before adding the additional category of missing [14]. The advantage of this approach was that only one model is fit and needed by the end-user. The missing indicator algorithm was similar to the categorization algorithm, but did not require categorization of continuous variables [15]. Instead, it introduced an indicator equal to 1 if the corresponding risk factor was missing versus 0 if not missing. The model included the indicator and the interaction with the risk factor. Since prostate volume was the only continuous risk factor that was sometimes missing, the missing indicator algorithm differed from the categorization algorithm in only one variable. Second-degree prostate cancer- and first-degree breast cancer family history were either both collected or not at all by the individual cohorts. Adding a missing category to them would therefore induce multi-collinearity. In order to avoid this, they were combined to a single new 5-category risk factor with second-degree prostate cancer family history only, first-degree breast cancer family history only, both present, none present or missing.

Multiple imputation has been recommended for fitting statistical models to training data to handle either outcomes or risk factors missing at random (MAR) [16]. In the case here, the outcome of clinically significant prostate cancer was not missing for any individuals so imputation was applied only for missing risk factors. Data were pooled across all ten cohorts to form the training set and imputation was applied using the pooled set and not by cohort. For a patient in the training set with multiple missing risk factors, multiple imputation by chained equations (MICE) sequentially imputes missing data according to full conditional models appropriate to the risk factor data type using all other risk factors available as covariates along with the outcomes that have been fit to complete cases in the training set [16, 17]. The R mice package uses 5 imputations as default and the literature has also recommended 10 iterations [16,18]. We implemented 30 imputations, as the average percentage of missing values across all risk factors in the training set, and averaged models built on the 30 imputed data sets for the final training set risk model. For the end-user or member of the validation set who is missing a risk factor, the algorithm imputed its value using mean values from the training set only, and not from other members of a validation set, as the latter would not be available in practice [17].

External validation on the European cohort, which was not used for training, was measured by discrimination using the area under the receiver-operating-characteristic curve (AUC) along with its 95% confidence interval (CI), calibration in the large (CIL), which evaluates the average difference between the predicted risk and binary clinically significant prostate cancer outcome for each patient in the validation set, and calibration-in-the-small by calibration curves of observed versus predicted risk according to deciles of predicted risk. Internal leave-one-cohort-out cross validation using the same metrics was also performed, by alternatively holding out one of the 10 PBCG cohorts used for training the model as a test set and training the models on the remaining 9 cohorts. Distributions of AUCs and CILs from the 10 test validations were visualized by violin plots showing smoothed histograms and boxplots showing medians and inter-quartile ranges. All analyses were performed in the R statistical package [19].

# 3. Results

Among 12,703 biopsies from 10 PBCG cohorts used for training, 3,597 (28%) had clinically significant prostate cancer, compared to 1,757 out of 5,540 (32%) clinically significant prostate cancer cases in the external validation cohort (Fig 1). All cohorts collected PSA and age but varied in collection of the other 10 risk factors, with some cohorts not collecting some risk factors at all (Fig 2). Differences between the cohorts in terms of distributions of the twelve risk factors and their associations with clinically significant prostate cancer are shown in Fig S1.

In leave-one-cohort-out internal cross-validation across the ten PBCG cohorts to ultimately be used for training the online model, the iterative BIC selection method had the lowest median CIL (-0.2%), while the available cases method had the highest (2.6%), all of which are minor in magnitude (Fig 3). CIL values ranged from -11 to 11% across the ten cohorts used as test sets. All six methods had nearly the same median AUC at 80%, and values ranged from 74 to 84% across the ten test sets. The categorization and missing indicator methods had larger variation in both the CIL and AUC than the other methods.

In external validation, all six methods either under- or over-predicted observed risks since none of the 95% CIs for CIL, computed as the average predicted risk minus the disease prevalence in the external validation cohort (32%), contained the value 0 (Table 2). The available cases method was the most accurate, under-predicting risk on average by 2.9%. The categorization and missing indicator methods over-predicted risks by 3.5% and 4.2%, respectively, while all other methods under-predicted risks, with imputation the worst by 12.4% (Table 2, Fig 4). The AUCs ranged from a low of 75.4% for the iterative BIC selection method to a high of 77.4% for the missing indicator method, but all 95% CIs overlapped (Table 2).

**Table 2.** External validation CIL and AUC values with risks as percentages along with 95% confidence intervals (CI).

| Method | CIL | (95% CI) | AUC | (95% CI) |
| --- | --- | --- | --- | --- |
| Available cases | -2.9 | (-4.0, -1.8) | 75.7 | (74.4, 77.1) |
| Iterative BIC selection | -8.6 | (-9.7, -7.5) | 75.4 | (74.0, 76.8) |
| Cohort ensemble | -7.1 | (-8.2, -6.0) | 76.4 | (75.1, 77.7) |
| Categorization | 3.5 | (2.4, 4.6) | 76.6 | (75.2, 77.9) |
| Missing indicator | 4.2 | (3.1, 5.3) | 77.4 | (76.1, 78.7) |
| Imputation | -13.3 | (-14.4, -12.2) | 75.9 | (74.5, 77.2) |

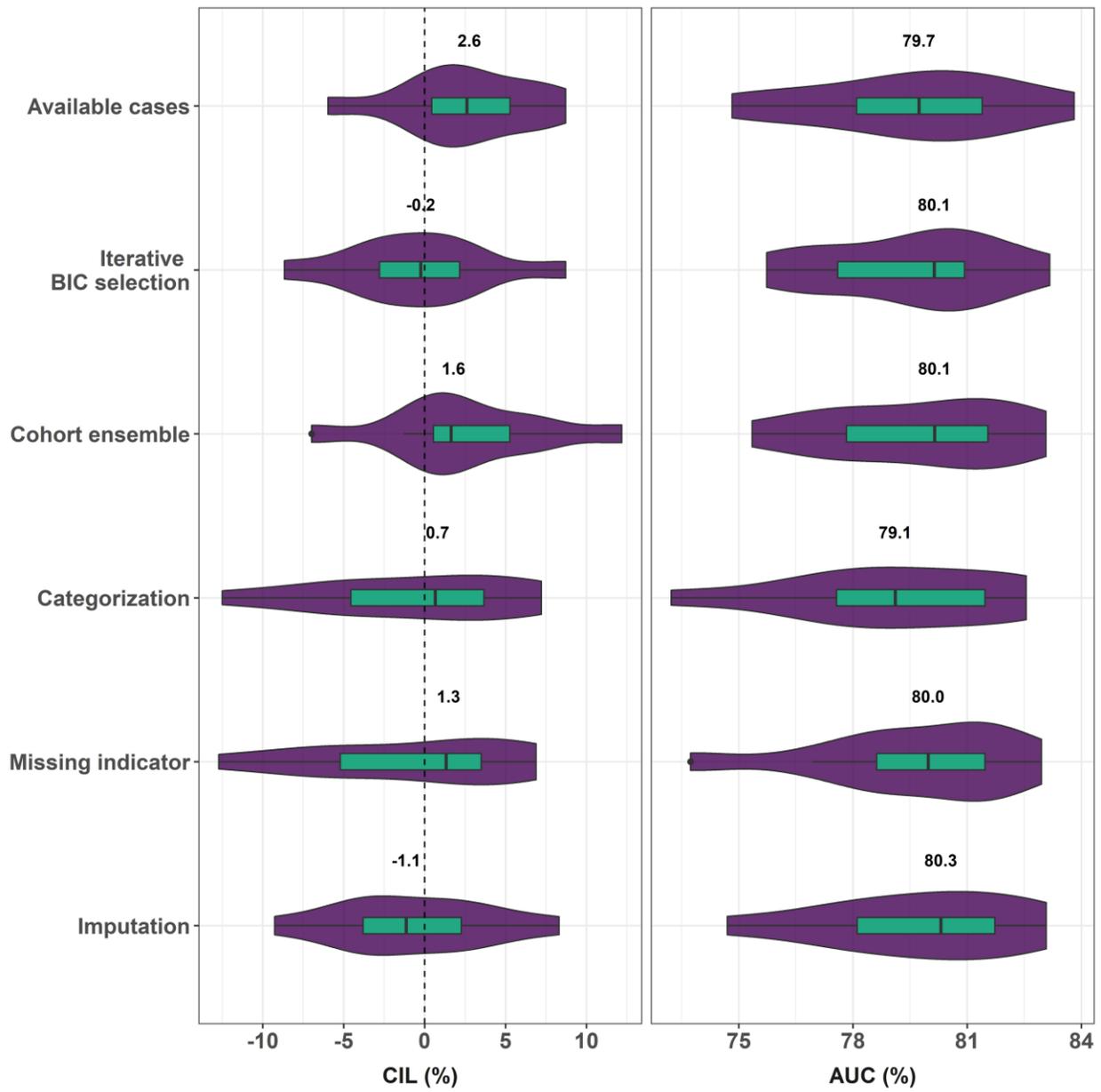

**Figure 3.** CIL and AUC performing leave-one-cohort-out cross-validation on 10 PBCG cohorts. Median values are indicated with numbers and as vertical lines in the boxes.

**Figure 4.** Calibration plots with shaded pointwise 95% confidence intervals for the 6 modeling methods applied to 10 PBCG training cohorts and validated on the external cohort. The diagonal black line is where predicted risks equal observed risks, lines below the diagonal indicate over-prediction, and lines above under-prediction, on the validation set.

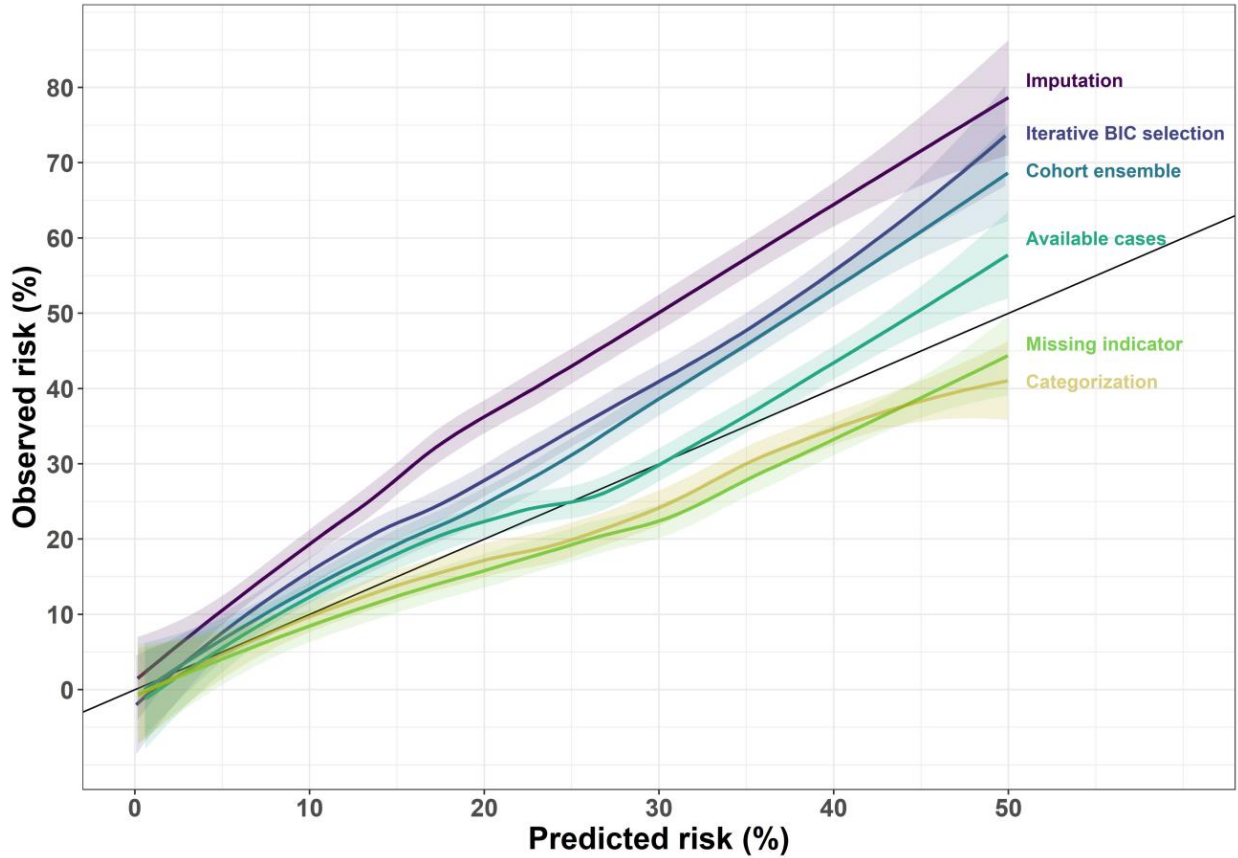

Comparisons of individual predictions from the six different methods for the 5,540 members of the external validation cohort are shown in Fig 5. As can be seen on the diagonal, for all methods the distribution of predicted risks for clinically significant prostate cancer cases were higher than for non-clinically significant prostate cancer individuals, but considerable overlap remained. Correlations of predictions by the 6 methods were high, all exceeding 0.8. The iterative BIC selection, cohort ensemble and available case methods were similar methods, all just using complete cases for the risk factor profile a specified individual has, and hence were highly correlated. The remaining three methods adjusted for missing data in some manner and were less correlated with these methods, with categorization the least correlated, though still very highly correlated.

We chose the available cases method for implementation of the risk tool online since it showed the most accuracy in terms of calibration in external validation (Fig 4), where all six methods showed equivalent AUCs (Table 2). AUCs and CILs across the 10 cohorts used as test sets in the internal leave-one-cohort-out cross-validation were also similar, and the available cases method had the lowest variability (Fig 3). The available cases method is less computationally intensive compared to multiple imputation and is valid under MAR assumptions based on unobserved risk factors and outcomes, which though untestable may be assumed as approximately holding when all established risk factors for outcomes are assumed to have been collected [20].

**Figure 5**. Marginal and pairwise comparisons of predictions from the 6 methods for the 5543 biopsies of the external validation set, pooled and stratified by clinically significant prostate cancer status (31.7% with clinically significant prostate cancer). Corr indicates Pearson correlation. Turquoise indicates individuals with clinically significant prostate cancer and purple not.

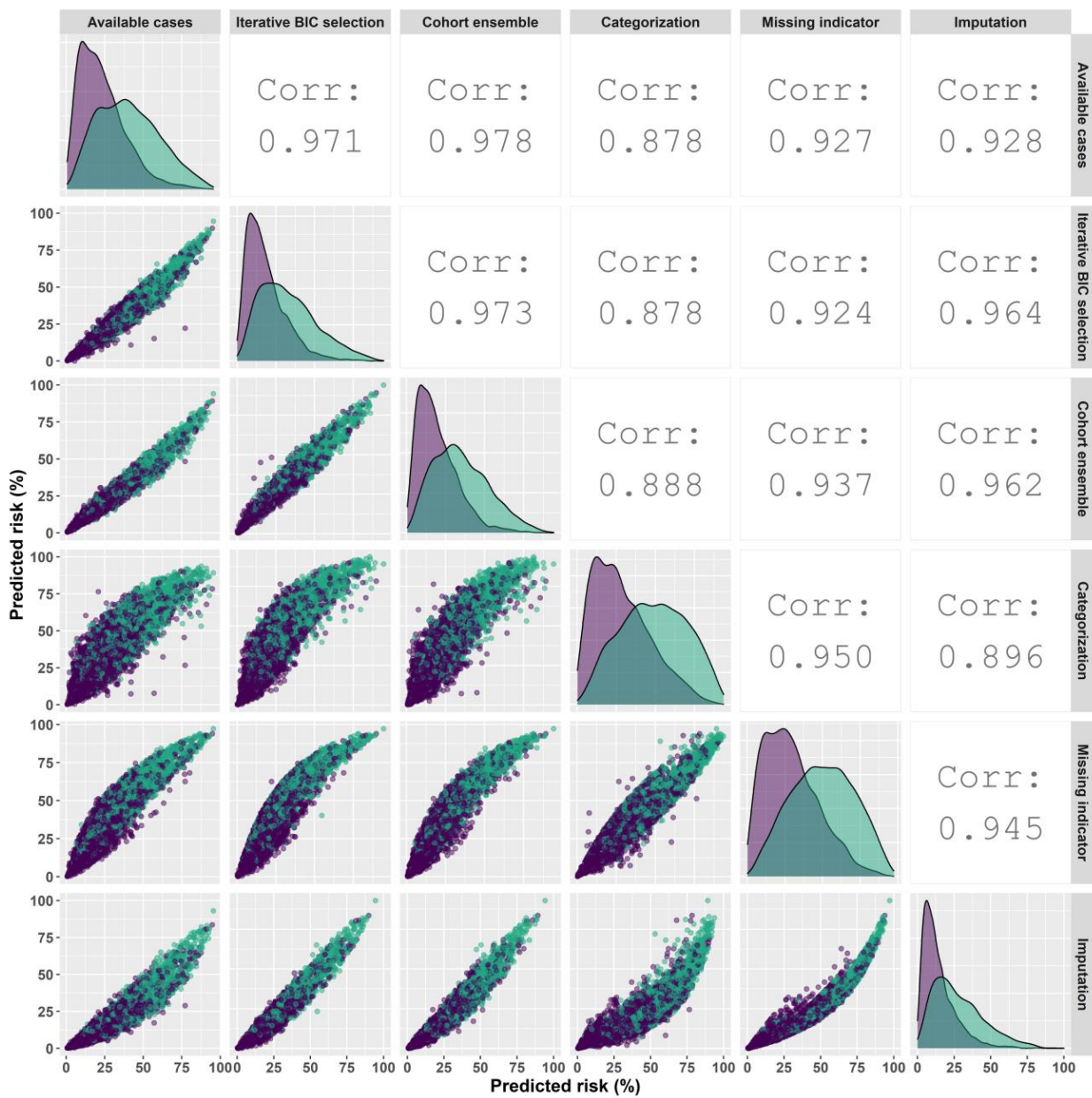

To implement the prediction tool online, we fit 1,024 models to allow for all possible missing risk factor patterns among 10 risk factors in order to use the maximum prostate biopsies possible from the 10 PBCG cohorts. The smallest model only contains PSA and age, utilizing all 12,703 biopsies from the 10 PBCG cohorts since these two risk factors were measured for all individuals. The largest model contains all 12 risk factors and was constructed from only 1,334 biopsies from 3 PBCG cohorts, as these were the only complete cases. These two risk models are shown in Table 3, with all possible models accessible online at riskcalc.org. Evaluated on the same validation set of 5,540 biopsies as used for Table 2, the original PBCG risk tool published in 2018 [2] based on only 6 of the 12 risk factors used here obtained a CIL of -5.9 (95% CI -7.1, -4.7), and an AUC of 66.9 (95% CI 65.4, 68.5), which is 10 points lower than any of the methods incorporating the additional risk factors. Adding just prostate volume to these six risk factors and evaluating on the validation set yielded a CIL of -10.1 (95% CI -11.2, -9.0), and an AUC of 75.6 (95% CI 74.2, 76.9), which are in line with the performance of all 12 risk factors in Table 2. Assessment of prostate volume, however, requires an invasive procedure that is not routinely performed in advance of the prostate biopsy.

**Table 3.** Odds ratios from the largest, standard, and smallest models in terms of number of 12 risk factors available from an end-user. Sample sizes are the number of individuals in the training set with all risk factors available (complete cases), and number of cohorts contributing the complete cases. In total 1,024 models are available based on the option for included versus not for 10 risk factors, all except PSA and age. R code for all 1024 models is available at the Cleveland Clinic Risk Calculator library, https://riskcalc.org/ExtendedPBCG/.

| Odds ratios for the full model containing 12 risk factors based on a fit to 1334 prostate biopsies from 3 cohorts. | | | |
|---|---|---|---|
| Risk factor | Odds ratio | 95% CI | p-value |
| Age | 1.07 | (1.05, 1.09) | < 0.0001 |
| PSA (log2) | 2.38 | (1.98, 2.89) | < 0.0001 |
| African ancestry | | | |
|     No | Ref | -- | -- |
|     Yes | 0.68 | (0.45, 1.03) | 0.08 |
| Prostate volume (log2) | 0.25 | (0.20, 0.32) | < 0.0001 |
| DRE | | | |
|     Normal | Ref | -- | -- |
|     Abnormal | 1.95 | (1.46, 2.60) | < 0.0001 |
| Prior negative biopsy | | | |
|     No | Ref | -- | -- |
|     Yes | 0.32 | (0.22, 0.45) | < 0.0001 |
| Hispanic ethnicity | | | |
|     No | Ref | -- | -- |
|     Yes | 1.08 | (0.78, 1.50) | 0.6 |
| 5-alpha-reductase-inhibitor use | | | |
|     No | Ref | -- | -- |
|     Yes | 0.96 | (0.63, 1.44) | 0.8 |
| Prior PSA screen | | | |
|     No | Ref | -- | -- |
|     Yes | 0.71 | (0.38, 1.34) | 0.3 |
| First-degree prostate cancer family history | | | |
|     No | Ref | -- | -- |
|     Yes | 1.93 | (1.38, 2.69) | 0.0001 |
| Second-degree prostate cancer family history | | | |
|     No | Ref | -- | -- |
|     Yes | 1.30 | (0.86, 1.96) | 0.2 |
| First-degree breast cancer family history | | | |
|     No | Ref | -- | -- |
|     Yes | 1.15 | (0.77, 1.70) | 0.5 |
| **Odds ratios for the model containing the 6 standard risk factors based on a fit to 8432 prostate biopsies from 9 cohorts.** | | | |
| Age | 1.05 | (1.04, 1.06) | < 0.0001 |
| PSA (log2) | 1.99 | (1.86, 2.12) | < 0.0001 |
| African ancestry | | | |
|     No | Ref | -- | -- |
|     Yes | 1.26 | (1.11, 1.44) | 0.0005 |
| DRE | | | |
|     Normal | Ref | -- | -- |
|     Abnormal | 2.57 | (2.29, 2.88) | < 0.0001 |
| Prior negative biopsy | | | |
|     No | Ref | -- | -- |
|     Yes | 0.28 | (0.24, 0.32) | < 0.0001 |
| First-degree prostate cancer family history | | | |
|     No | Ref | -- | -- |
|     Yes | 1.94 | (1.70, 2.22) | < 0.0001 |
| **Odds ratios for the smallest model containing 2 risk factors based on a fit to 12703 prostate biopsies from 10 cohorts.** | | | |
| Age | 1.05 | (1.05, 1.06) | < 0.0001 |
| PSA (log2) | 1.72 | (1.64, 1.80) | < 0.0001 |

# 4. Discussion

Systematic missing clinical data across heterogeneous cohorts poses analysis challenges for both model developers and end-users. We compared six methods that have been proposed for handling missing data, ultimately choosing the pooled available case method. The available case method has been recommended by statisticians as it is valid even when data are MAR, in this context, meaning that whether or not a risk factor is missing is independent of the unknown value of the risk factor conditional on the known outcome and other available risk factors for the patient [20, 10, 21]. The majority of risk factors collected across the PBCG are those typically collected in urological clinics from men presenting for PSA screening or follow-up. The most ubiquitous and predictive risk factors, PSA and age, have been collected for all PBCG participants, and so are exempt from missing data assumptions. The assumption of MAR for the remaining risk factors may be questionable in some cases, for example prostate volume may not have been reported when the value was assessed to be too low or clinically significant prostate cancer was not discovered on biopsy. There is no statistical test for MAR, hence we relied on external and internal cohort-based validations to compare the available case to competing methods for selection of the method producing optimal performance across a range of scenarios that would be encountered in practice.

The missing indicator method has been shown to potentially result in biased odds ratios, even when data are missing completely at random, meaning no relation, conditional or not, between whether a risk factor is missing and all other variables, leading to strong recommendations against its use for causal or explanatory inference [20, 10, 22]. The categorization method suffers from the same potential biases since it changes all continuous predictors to categorical ones before applying the missing indicator method. A recent study affirmed that such methods could be used for randomized trials as the missing-ness of protocol-specified variables would be randomized by the random treatment assignment, thus eliminating systematic bias [15].

The emergence of clinical risk prediction tools embedded in electronic health records, where missing data are large and systematic, has led to support for the missing indicator method used in model development to match the method used when the model is deployed, and that if informative presence is potentially informative with respect to prediction, then it should be leveraged [23, 24]. Machine learning and other supervised learning methods follow the principle of developing prediction models to optimize accuracy on internal and external validation, often with uninterpretable models. The renown James-Stein result shows that an estimator with effects shrunk towards zero can be preferable to the unbiased estimator, and these concepts are often applied in regularized regression approaches for situations with high numbers of predictors [25]. Across the validations performed in the PBCG, the potentially biased missing indicator and categorization methods did not perform substantially worse than the available cases methods. But we agree that caution should be exercised towards their use when data are combined across cohorts, where some cohorts do not collect some risk factors at all, as was the case with extended family history in this study. In this case, the effect of the missing category was confounded with that of cohort. The coefficient for missing prostate volume following the missing indicator method fit to the 10 PBCG cohort data was close to zero, meaning a patient with missing prostate volume had nearly 0 odds of clinically significant prostate cancer compared to a person not missing prostate volume, which can only be a cohort effect.

In addition to contributing to model development techniques for systematic missing data across heterogeneous cohorts, we have provided helpful methods for the end-user of online risk tools, namely the fit of multiple models for different risk factor missing data patterns. Such work enables more users to access online risk tools. Each model was fit to all complete cases that contained the risk factors, thus optimizing information and accuracy for the user. Our online tool requires PSA and age for use, and any collection of up to 10 additional risk factors. As consortia and available data grow in size, so does the amount of missing data. A flexible modeling strategy accommodating missing data on both the development- and user-end maximizes information by utilizing multiple data sources and increases accessibility to a broader band of patients, by including those limited in risk factor assessment.

# Supporting information

**Table S1.** Algorithms for the 6 risk modeling approaches. Starting variables are available risk factors from the user. Used and cohort variables are subsets or all of the starting variables corresponding to those used by the model and those with less than 40% missing rates in the cohort, respectively.

| | Algorithm: Available cases |
|---|---|
| 1 | Subset the PBCG dataset by records without missing values for *starting variables* |
| 2 | Fit main effects logistic regression model with *starting variables* to the PBCG subset pooled for all cohorts |
| | **Algorithm: Iterative BIC selection** |
| 1 | *used variables = starting variables* |
| 2 | while number of *used variables* reduces do |
| 3 |     Subset the PBCG dataset (pooled for all cohorts) by *used variables* |
| 4 |     Use only complete records |
| 5 |     Perform logistic regression BIC selection starting with main effects up to two-way interactions |
| 6 |     *used variables* = variables in the selected model (either as main effect or as interaction) |
| 7 |     *risk* = Predict clinically significant prostate cancer risk with information in *starting variables* |
| 8 | end while |
| | **Algorithm: Cohort ensemble** |
| 1 | *cohort variables* = variables of each cohort with less than 40% missing records |
| 2 | for all cohorts do |
| 3 |     *used variables* = variables of *starting variables* that are in *cohort variables* |
| 4 |     Subset the PBCG dataset by cohort, and *used variables* |
| 5 |     Use only complete records |
| 6 |     Perform logistic regression BIC selection starting with main effects up to two-way interactions |
| 7 |     *used variables*[cohort] = variables in the selected model (either as main effect or as interaction) |
| 8 |     *risk*[cohort] = Predict clinically significant prostate cancer risk with information in *starting variables* |
| 9 | end for |
| 10 | *overall risk* = mean(*risk*) |
| | **Algorithm: Categorization** |
| 1 | Categorize all predictor variables with the additional factor not available (NA). Continuous variables with missing values are categorized. |
| 2 | Fit main effects logistic regression model with all variables to the PBCG dataset pooled for all cohorts |
| **Algorithm: Missing indicator** | |
| 1 | Categorize all categorical predictor variables with the additional factor missing. Add for continuous variables with missing values an indicator variable whether the variable was missing or not |
| 2 | Fit main effects logistic regression model with all variables and include the interactions of the indicator with the corresponding variable in the model to the PBCG dataset pooled for all cohorts |
| **Algorithm: Imputation** | |
| 1 | Run 30 imputations by chained equations (mice) on the training PBCG dataset |
| 2 | for all imputed datasets do |
| 3 |     Fit main effects logistic regression model with all variables in the model to the imputed PBCG dataset pooled for all cohorts |
| 4 | end for |
| 5 | Impute missing values in the test case with means of the training set |
| 6 | Average the coefficients from mice for use in the prediction models |

**Figure S1**. Differences between the cohorts in terms of distributions of the twelve risk factors and their associations with clinically significant prostate cancer. NA = missing.

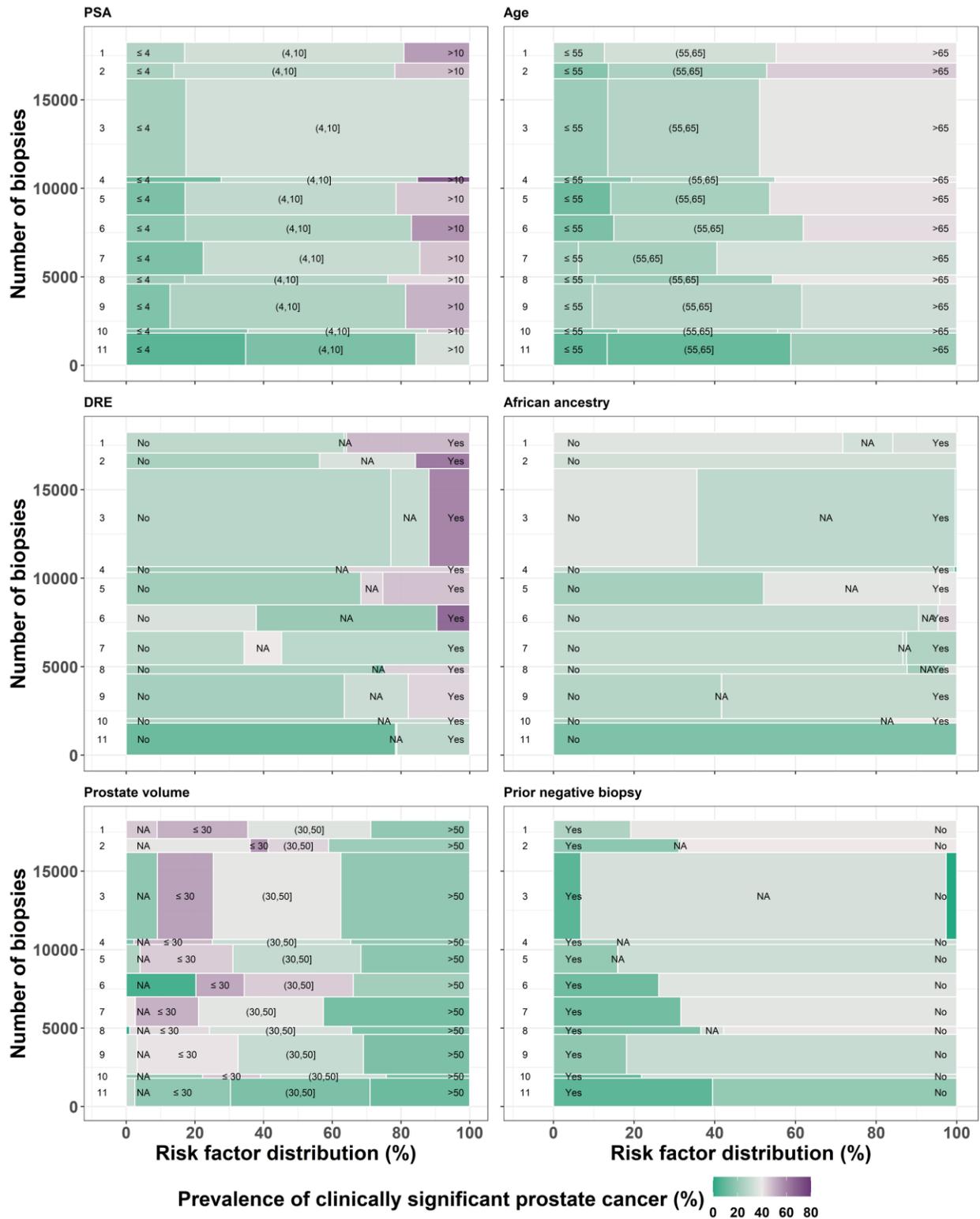

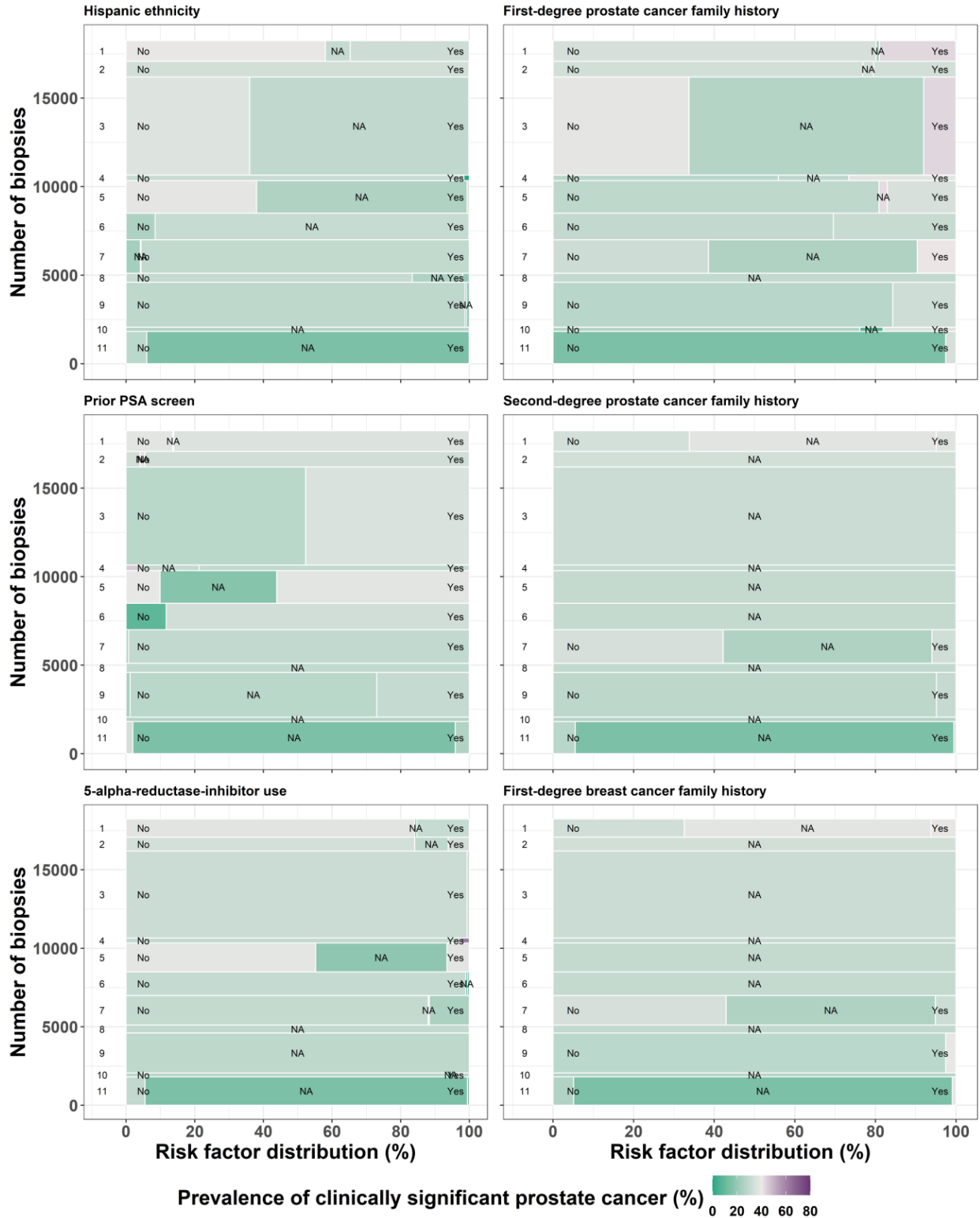